\documentclass{aa}  
\usepackage{graphicx}
\usepackage{txfonts}
\usepackage[colorlinks,allcolors=blue]{hyperref}

\begin{document} 


	  \title{Numerical simulations of the 1840s great eruption of $\eta$ Carinae}

  \subtitle{I. Revisiting the explosion scenario}

   \author{R. F. Gonz\'alez\inst{1}\and Luis A. Zapata\inst{1}\and  A. C. Raga \inst{2}\and
 J. Cant\'o \inst{3}\and P. F. Vel\'azquez \inst{2}\and E. M. de Gouveia Dal Pino \inst{4}
          }

   \institute{Instituto de Radioastronom\'ia y Astrof\'isica (UNAM),
   			A.P. 3-72 C.P. 58190 Morelia, Michoac\'an, M\'exico \\
              \email{rf.gonzalez@irya.unam.mx;l.zapata@irya.unam.mx}
		\and
			Instituto de Ciencias Nucleares (UNAM), A.P. 70-543,
			 C.P. 04510, M\'exico, M\'exico \\
 			 \email{pablo@nucleares.unam.mx; raga@nucleares.unam.mx}
			 \and
			 Instituto de Astronom\'ia (UNAM), A.P. 70-264, C.P. 04510,
			 M\'exico, M\'exico \email{jcanto@astro.unam.mx}
			 \and
			 Instituto de Astronomia, Geof\'isica e Ci\^encias Atmosf\'ericas
			 (IAG-USP), Rua do Mat\~ao, S\~ao Paulo 05508-090, Brazil)
             }

   \date{Received ; accepted }

 
  \abstract{
In this work, we present new two-dimensional hydrodynamical simulations of the
major eruption of $\eta$ Car in the 1840s, which resulted in the formation
of a bipolar nebula that is commonly known as the {\it large} Homunculus.
In our numerical models, we have included the high-speed component of 10000 km
s$^{-1}$, detected in recent observations, that provides direct evidence of an
explosive event. Here, we investigate whether such a violent explosion
is able to explain both the shape and the dynamical evolution of $\eta$ Car's
nebula. As in our previous work, we have assumed a two-stage scenario for
$\eta$ Car's eruption: a slow outflow phase during a few decades before the
eruption followed by the explosive event. From the collision of these outflow
phases, the {\it large} Homunculus is produced. Our numerical
simulations show that such scenario does not resemble some of the observed
physical features and the expansion of the nebula. Notwithstanding, we also
explore other injection parameters (mass-loss rate and ejection velocity) for these
outflow phases. In particular, we find that an explosion with an intermediate-speed
of 1000 km s$^{-1}$ is able to reproduce the morphology and the kinematical age of
the {\it large} Homunculus.
	}


   \maketitle
%

\section{Introduction}

The origin of the 19th century major eruption of the massive star $\eta$
Car is still in an intense debate \citep[{e.g.}][]{por2016,smi2006}. 
The $\eta$ Car system is a bright (4 $\times$ 10$^{6}$ L$_{\odot}$)
and massive stellar system that contains an evolved (Luminous Blue Variable
LBV) star with a mass of $\sim$ 90 M$_{\odot}$ and a less massive ($\sim$ 30
M$_{\odot}$) main sequence star companion. The orbital period of such an
eccentric system ($e = 0.9$) is only 5.5 yr \citep{por2016}. The primary
star of $\eta$ Car's system has undergone several eruptive events, being
the most notorious the major eruption in the 1840s  \citep{dav1997} 
during which $\eta$ Car drastically increased its visual
magnitude \citep{hum1999}, and expelled a large amount of mass
($\ge$ 10 M$_{\odot}$) into its surroundings \citep{smi2003}.
From this eruption, a bipolar nebula known as the {\it large} Homunculus was
formed which extends $\pm$ 9 arcsec along the symmetry axis (that corresponds
to a physical size of $\pm$ 3.25 $\times$ 10$^{17}$ cm at a distance of 2.3 Kpc)
and expands at a speed of $\sim$ 650 km s$^{-1}$ in the polar direction
\citep[see,][]{smi2006}. Its symmetry axis is inclined $\sim$ 45$^{\circ}$
to the line of sight \citep{dav2001}. In addition, there were at least two
other outbursts after this event in the years of 1890s and 1940s.
From these eruptions the so-called {\it little} and {\it baby} Homunculi were
produced, respectively \citep{ish2003,abr2014}.

\begin{table*}
\begin{center}
\leftline{{\bf Table 1}.  Models of $\eta$ Car's Homunculi: 
Parameters of the colliding-outflows.}
\vskip 0.2cm
\begin{tabular}{llccccrrrrrrc}
\hline
\hline
\noalign{\smallskip}
 & & Outflow Phase & & $\lambda$ & & $v_1$ [km s$^{-1}$] & & $v_2$
 [km s$^{-1}$] & & $\Delta t$ [yr] & & $\dot m$ [M$_{\odot}$ yr$^{-1}$]\\
\hline
Run A & & Standard wind & & 2.4 & & 250 & & 14 & & 507  & & 1.0 $\times $10$^{-3}$ \\
& & Pre-outburst wind & & 1.9 & & 500 & & 14 & & 20 & & 0.7 \\
& & Explosion & & 1.9 & & 1000 & & 100 & & 1 & & 1.0 \\
& & Post-outburst wind & & 1.9 & & 500 & & 14 & & 49 & & 1.0 $\times $10$^{-3}$ \\
& & Minor Eruption & & 1.9 & & 200 & & 10 & & 10 & & 1.0 $\times $10$^{-2}$ \\
& & Post-outburst wind & & 1.9 & & 500 & & 300 & & 120 & & 1.0 $\times $10$^{-3}$ \\
\noalign{\smallskip}
\hline
Run B & & Standard wind & & 2.4 & & 250 & & 14 & & 507  & & 1.0 $\times $10$^{-3}$ \\
& & Pre-outburst wind & & 1.9 & & 500 & & 30 & & 20 & & 0.7 \\
& & Explosion & & 1.9 & & 10000 & & 100 & & 1 & & 0.1 \\
& & Post-outburst wind & & 1.9 & & 500 & & 14 & & 49 & & 1.0 $\times $10$^{-3}$ \\
& & Minor Eruption & & 1.9 & & 200 & & 10 & & 10 & & 1.0 $\times $10$^{-2}$ \\
& & Post-outburst wind & & 1.9 & & 500 & & 300 & & 120 & & 1.0 $\times $10$^{-3}$ \\
\noalign{\smallskip}
\hline
\end{tabular}
\begin{flushleft}
{\bf Notes.} The {\it 1890s minor eruption} and {\it post-eruptive wind}
conditions are taken from \citet{gon2010}.
\end{flushleft}
\end{center}
\end{table*}

The most successful explanation of the nature and the origin of the
Homunculus nebula is given by the stellar merger model \citep{por2016,smi2018}. 
The merger between two massive stars in a triple hierarchical system could account
for most of the physical characteristics observed in $\eta$ Car.
These physical features include the kinetic energy (10$^{49.6}$ -
10$^{50}$ erg), the luminosity burst (4 $\times 10^{6}$ L$_{\odot}$),
the bipolar shape of the Homunculus, and the two successive stages of
the velocity field during the main eruption: one very broad $\sim$
10000 km s$^{-1}$, and the other one of  $\sim$ 600 km s$^{-1}$
\citep{smi2018}. Nonetheless, there are some other models for explaining
the nature of the {\it large} Homunculus, such as super-Eddington
eruptions  \citep{owo2016}, binary interactions at periastron
\citep{kas2009}, and pulsational pair instabilities in a massive
star \citep{woo2017}, but they do not account for all the observed features
in the nebula.

The violent eruption of the mid-nineteenth century suffered by $\eta$
Car was analogous to a type II supernova explosion, but with lower
energy released to the interstellar medium (ISM) (E$_{k} \sim 10^{50}$ erg
over a few years) and therefore, it is referred to as a supernova
impostor \citep{smi2013}. The extremely broad emission wings of about
$\sim$ 10000 km s$^{-1}$ in the H$_{\alpha}$ optical lines are likely caused
by high-velocity outflowing material occurred during the explosion
\citep{smi2018b}. In a recent paper, \citet{gon2018}  investigated through
two-dimensional hydrodynamical simulations whether an explosive event is able to
explain the shape and kinematics of the {\it large} Homunculus nebula.
These numerical models revealed that in fact the explosion can indeed
explain some observed features in the nebula, such as the present-day
double-shell structure, and thermal instabilities (Rayleigh-Taylor
and Kelvin-Helmholtz) along the dense shell. Nevertheless, the explosion
scenario proposed by \citet{smi2003} could not account for the current
physical size of the {\it large} Homunculus and, consequently, the
estimated age of the nebula. It is noteworthy that these simulations
do not include the high-velocity component of $\sim$ 10000 km s$^{-1}$
reported more recently by \citet{smi2018,smi2018b}. In this work,
we revisit 2D numerical simulations of the major eruption of $\eta$
Car in the 1840s adopting the explosion scenario proposed by \citet{smi2013}, 
including in our models this component of the outflowing
gas.

The paper is organized as follows. In $\S$ 2, we describe the model.
The numerical simulations and discussion of the results are presented
in $\S$ 3. Finally, in $\S$ 4 we draw our conclusions.

\section{The model}

In this work, we assume the two-stage scenario for $\eta$ Car’s eruption
proposed by \citet{smi2018}: (1) a slow and massive wind phase expelled for
a few decades before the 1840s, and (2) a lighter and faster explosive event.
We have carried out two distinct models. In Run A, we have adopted an
expansion speed of 1000 km s$^{-1}$ for Stage 2, while in Run B, an extremely
high-velocity of 10000 km s$^{-1}$ is assumed for the explosion. According to
\citep{smi2018} from the circumstellar medium (CSM) interaction between the
two stages, a dense and fast shell is produced that became the {\it large}
Homunculus. Here, we explore such scenario and investigate whether it is able
to explain not only the shape of the nebula, but also some of its observed
physical properties, such as the latitude-dependent expansion speed and the
estimated dynamical age.
It is noteworthy that the formation of the internal nebula (commonly known
as the {\it little} Homunculus) as well as the high-speed features in the
equatorial skirt of $\eta$ Car have been investigated in our previous works
\citep{gon2004,gon2004b,gon2010}.

For the different outflow phases, we have assumed the latitude-dependent
ejection velocity and density proposed in Gonz\'alez et al. (2010), that is,

\begin{eqnarray}
v = v_1\, F(\theta) \, ,
\end{eqnarray}

\noindent
and,

\begin{eqnarray}
n = n_0\, {\biggl({{r_0}\over{r}}\biggr)}^2 \, {{1}\over{F(\theta)}} \, ,
\end{eqnarray}

\noindent
respectively, where $r_0$ is the injection radius, and $F(\theta) =
[(v_2/v_1) + e^{2z}]/[1 + e^{2z}]$ being $z = \lambda$ cos$(2\theta)$ a
function that controls the shape of the {\large} Homunculus.
 Equation (1) is a parametrization for the velocity of a latitude-dependent
wind which that was first proposed by \citet{ick1988}, and has been extensively
used for modelling aspherical wind bubbles. The constant
$\lambda$ is related to the degree of asymmetry of the outflow,
and $\theta$ is the polar angle. The speed $v_1$ is related to the expansion
velocity $v_p$ in the polar direction ($\theta = 0^{\circ}$), while $v_2$
to the corresponding value $v_e$ at equator ($\theta = 90^{\circ}$).
From equations (1) and (2), it follows that a
constant mass-loss rate per unit solid angle is assumed. It is worth
mentioning that \citet{gon2010} found the best fit to the observed
expansion speed at different latitudes of $\eta$ Car’s Homunculus \citep{smi2006} 
using $\lambda = 1.9$, $v_1$ = 670 km s$^{-1}$, $v_2$ = 100
km s$^{-1}$. Using these parameters, the .predicted expansion velocities
are $v_p$ = 657.5 km s$^{-1}$ at the poles and $v_e$ = 112.5 km s$^{-1}$
in the equatorial direction \citep[see Fig. 1 of ][]{gon2010}.

\begin{figure*}
\hspace{0.4cm}\includegraphics[width=9.0in,angle=0]{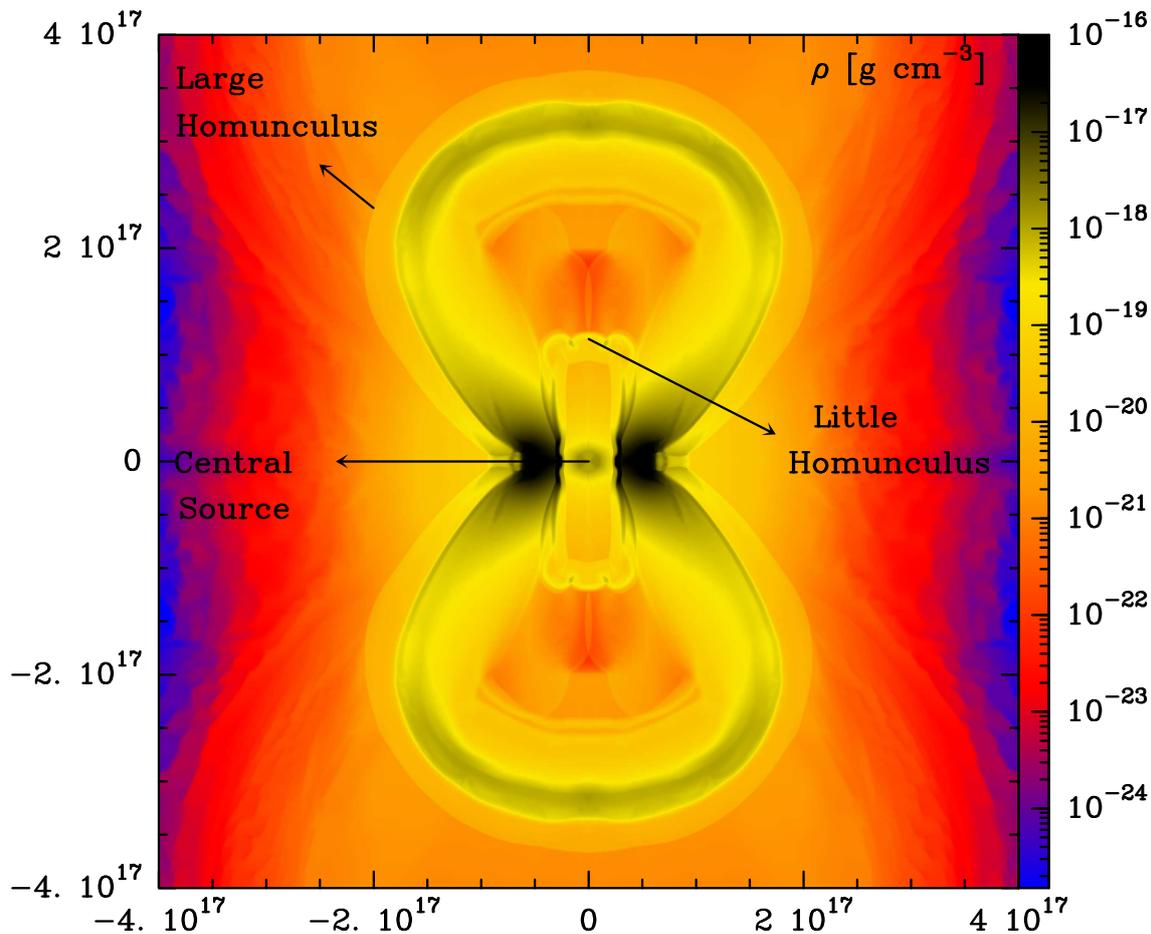}\hspace*{\fill}
\vspace{-3cm}
    \caption{Log-scale density stratification of $\eta$ Car's nebulae.
The numerical simulation (Run A) corresponds to an evolution time of $t =$ 172 yr
since the great eruption in the 1840's. The bar at the right-side of the plot is
in units of g cm$^{-3}$, and the computational domain axes are in units of cm.
See the text for a physical description of the figure.}
    \label{fig1}
\end{figure*}

\section{Numerical simulations}

We performed new gas dynamic 2D numerical simulations of $\eta$ Car’s
major eruption of the 1840s. The simulations were carried out
with the adaptive grid Yguaz\'u-A code developed by \citet{rag2000}.
 This code integrates the continuity, momentum and energy equations
written in "conservation form". Also, the adaptive grid points are
created using linear interpolations, which preserve the energy, mass
and the momenta along the three coordinate axes. The problem that we
are treating, though, is highly non-conservative due to the presence
of a strong energy loss term. Different tests of the code have been
reported by \citet{rag2000, rag2001,vel2001}.

The version of the code that has been used
\citep{gon2018} solves the gas dynamic equations explicitly
accounting for the radiative cooling for the atomic/ionic species HI, HII,
HeI, HeII, HeIII, CII, CIII, CIV, OI, OII, OIII, OIV. The cooling rates
  for the individual ions were calculated with the analytic approximations
  of Raga et al. (2002), which include the collisionally excited lines and
  collisional ionization (by free electros), and a parametrized cooling
  rate for high temperatures. A set of rate equations is integrated together
  with the gas-dynamic equations to compute the non-equilibrium ionization
  state (with the ions listed above).
  The abundances (by number, relative to the total
  number of atoms) are of 0.9 (H), 0.1 (He), $6.6\times 10^{-4}$ (C)
  and $3.3\times 10^{-4}$ (O).

The calculations were
axisymmetric and performed using the flux-vector splitting algorithm of
\citet{van1982} on a five-level binary adaptive grid with a maximum resolution
of 3.9 $\times$ 10$^{14}$ cm along the two axes. The computational domain
extends over (4 $\times$ 10$^{17}$) $\times$ (4 $\times$ 10$^{17}$), which is
initially filled by a homogeneous medium of density of 10$^{-3}$ cm$^{-3}$ and
temperature of 10$^{2}$ K.

We should note that in our simulations we have assumed
optically thin cooling, but the inner region of the Eta Carinae
nebula are likely to be at least partially optically thick. We therefore
overestimate the cooling rate (and therefore obtain lower temperatures)
in the inner regions of the computed flow.

In Table 1, the models (Runs A-B) performed for $\eta$ Car's
major eruption are listed. In these models, the adopted parameters
of the minor eruption and the winds expelled before and after this
eruptive event are taken from \citet{gon2018}. In addition,
the physical conditions for the pre-eruptive wind and the explosion
are consistent with the two-stage shock-powered event proposed by
\citet{smi2018} for $\eta$ Car's major eruption. In our numerical
models, a standard wind with terminal velocity of $v_0 = 250$ km
s$^{-1}$ (in the polar direction) and mass-loss rate of $\dot M_0 =$ 10$^{-3}$
M$_{\odot}$ yr$^{-1}$ is blown from a distance of $r_0$ = 10$^{16}$
cm with a temperature $T_0 =$ 10$^5$ K into the unperturbed environment.
Once the computational domain is filled by this wind, the double-stage
of the major eruption occurs.

In Run A, the slow pre-outburst wind with an ejection velocity of 500
km s$^{-1}$ (at the poles) and mass-loss rate of $\dot M =$ 0.7 M$_{\odot}$
yr$^{-1}$ is ejected during 20 yr, that results in a total mass of 14
M$_{\odot}$ expelled during this stage; and a brief explosion that
lasts 1 yr, in which a mass of 1 M$_{\odot}$ is released with a speed
of $v = 1000$ km s$^{-1}$ from the central source. 
Accordingly, the total kinetic energy released during the explosion
is $\sim$ 10$^{49}$ erg, which is consistent with the estimated value
of 10$^{49.7}$ by \citet{smi2013}. 
Afterwards, a {\it post-outburst} wind with similar conditions of
the current wind ($v = 500$ km s$^{-1}$ and $\dot M_0 =$ 10$^{-3}$
M$_{\odot}$ yr$^{-1}$) is ejected until the minor eruption occurs,
when the ejection velocity drops (200 km s$^{-1}$) and the mass-loss
rate increases ($\dot M_0 =$ 10$^{-2}$ M$_{\odot}$ yr$^{-1}$). Later,
a post-eruption wind with $v = 500$ km s$^{-1}$ and $\dot M_0 =$
10$^{-3}$ again resumes, and the {\it little}
Homunculus is formed \citep{ish2003}. In our previous
papers \citep{gon2004,gon2010} the
formation and dynamical evolution of the internal nebula has been
investigated, and consequently, we focus here on the physical
properties of the {\it large} Homunculus only.

\begin{figure*}
\hspace{-2.0cm}\includegraphics[width=10.5in,angle=0]{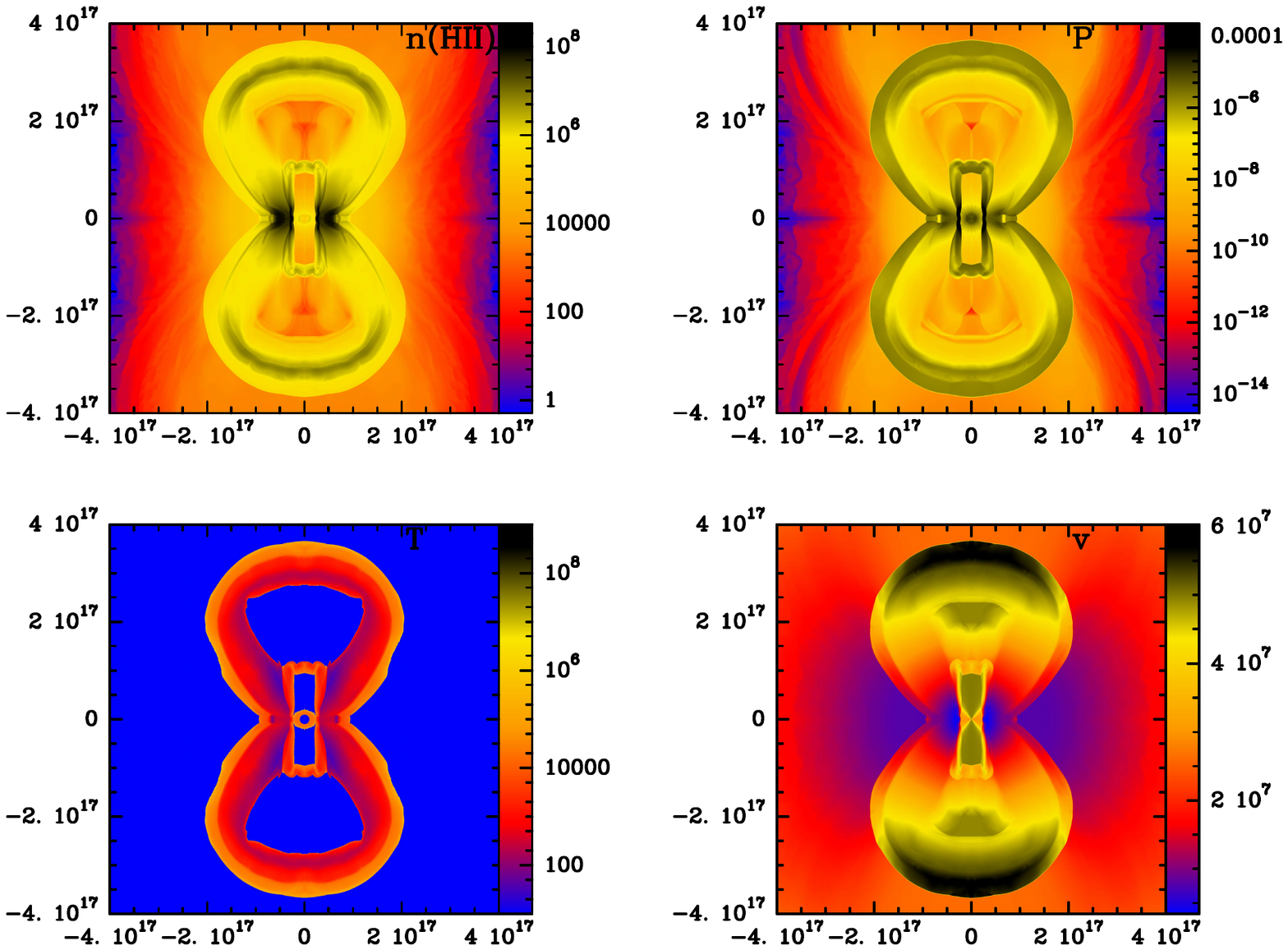}
\vspace{-4cm}
    \caption{Stratifications for Run A of density of ionized hydrogen (top left), pressure
(top right), temperature (bottom left) and velocity (bottom right) computed 172
yr after the great eruption of $\eta$ Car. The bars at the right-side of the panels
are in units of cm$^{-3}$, dyn cm$^{-2}$, K, and cm s$^{-1}$, respectively, and
the computational domain axes are in units of cm. A further description of the
figure is given in the text.}
    \label{fig2}
\end{figure*}

\begin{figure*}
\hspace{0.4cm}\includegraphics[width=9.0in,angle=0]{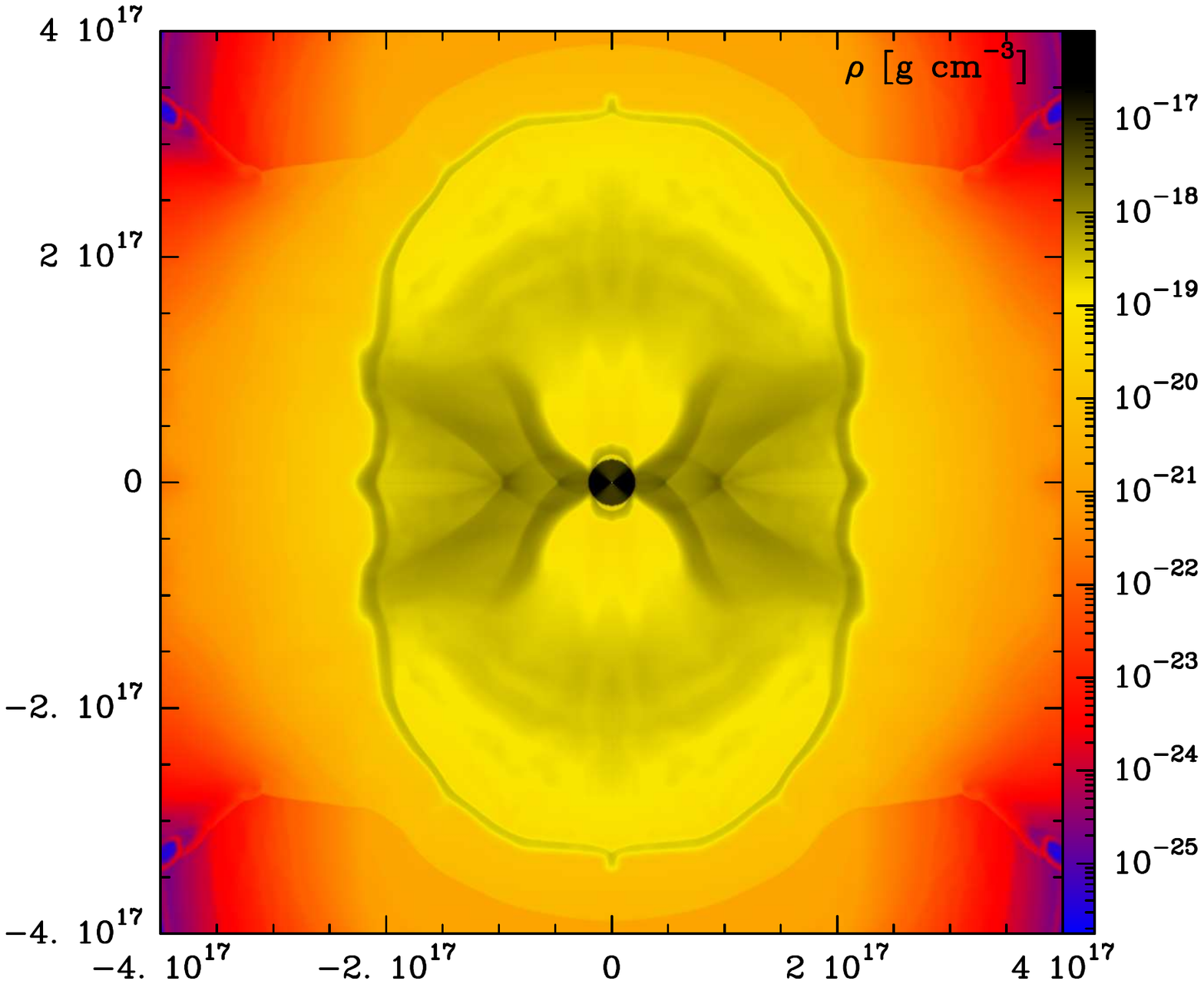}\hspace*{\fill}
\vspace{-3cm}
    \caption{Same as figure 1 of $\eta$ Car's nebulae for Run B after
$t =$ 52 yr since the explosion. The bar at the right-side of the plot is
in units of g cm$^{-3}$, and the computational domain axes are in units of cm.
A physical description of the figure is given in the text.}
    \label{fig3}
\end{figure*}

Figure 1 depicts the predicted density distribution obtained from Run A
of $\eta$ Car's nebulae . The log-scale map at time $t =$ 172 yr after
the major eruption is presented. It can be seen from the figure
that the interaction of the different outflow phases produces
internal and external nebulae which are consistent with the physical
size and the shape of both Homunculi. For the large Homunculus we obtain
a physical extension of the polar lobes of $\pm$ 3.5 $\times$ 10$^{17}$ cm,
expanding at a speed of $\simeq$ 600 km s$^{-1}$, which is close to the
observed value of $\simeq$ 650 km s$^{-1}$ of the
Homunculus at high latitudes \citep{smi2006}. At the equator, the predicted
physical size of the nebula across the star is $\pm$ 9.1 $\times$ 10$^{16}$
cm expanding at $\simeq$ 140 km s$^{-1}$. In addition, as predicted
in our previous work \citep{gon2010}, the inner Homunculus
shows the formation of instabilities (Rayleigh-Taylor and
Kelvin-Helmholtz) along the polar caps due to the interaction of the
lower density post-outburst wind that pushes and accelerates a higher
density outflow (the 1890's minor eruption). After 172 yr of evolution,
they expand at $\simeq$ 350 km s$^{-1}$ extending from - 1.1 $\times$
10$^{17}$ cm to + 1.1 $\times$ 10$^{17}$ cm along the major axis.
It is worth mentioning that in \citet{gon2004,gon2010}, 
we showed that the interaction between both Homunculi
in the equatorial zone results in the formation of tenuous, high-speed
equatorial features ($\sim$ 1000 km $s^{-1}$) that seem to be related
to the observed equatorial skirt of $\eta$ Car nebulae.
Notwithstanding, we note that these features are not
produced in this numerical model, probably because the massive pre-outburst
wind has a significant effect on such interaction. We have assumed in
the simulation that 14 M$_{\odot}$, of a total of 15 M$_{\odot}$ expelled
during the two-stage scenario proposed by \citep{smi2018} for $\eta$ Car's
eruption, are ejected during the pre-outburst wind. In contrast, the
numerical models presented in \citet{gon2018} assumed 10 M$_{\odot}$ for
each stage, which allows for the formation of the high-speed material at equator
of $\eta$ Car's lobes.

In Figure 2, we present the stratifications for Run A of the density of
ionized hydrogen (top left), pressure (top right), temperature (bottom left)
and velocity (bottom right) computed at an evolution time of $t =$ 172 yr after
of the explosion phase.
As we mentioned above, the simulation shows the formation of an external and
internal nebulae which are in good agreement with the shape and kinematics of
the large and the little Homunculi. The simulation shows a temperature
of $T \simeq 10^4$ K behind the outer shock,
while the mean temperature inside the shocked shell is $T \simeq$ 500 K.
In addition, it can be observed a low-density cavity between both Homunculi
with temperature of $T \simeq$ 10 K. On the other hand, no equatorial
high-speed, low-density features that may be related to the observed equatorial
skirt of $\eta$ Car nebula are produced.

In Run B, we also assume a slow pre-outburst wind ejected for 20 yr
with an ejection velocity of 500 km s$^{-1}$ (at the poles) and mass-loss
rate of $\dot M =$ 0.7 M$_{\odot}$ yr$^{-1}$. Afterwards, a brief
explosion that lasts 1 yr occurs, during which a total
mass of 0.1 M$_{\odot}$ is expelled from the central source with a speed
of $v = 10000$ km s$^{-1}$ (in the polar direction). It is worth mentioning
that in this model we suppose for the explosive event a latitude-dependent
mass-loss rate $\dot M \propto  F(\theta)$, so the density $n$ is not a
function of $\theta$ during this stage (see $\S$ 2). Adopting these
parameters, the total kinetic energy released during the explosion is
$\sim$ 10$^{50}$ erg, which is very close to the estimated value of
10$^{49.7}$ by \citet{smi2013} for the great eruption of $\eta$ Car in
the 1840s. The physical conditions for the subsequent outflow phases are
the same as Run A. Accordingly, this numerical simulation includes the
high-speed component of 10000-20000 km s$^{-1}$ for the explosion stage
reported recently by \citet{smi2018,smi2018b}. In addition, it is also
remarkable that, as pointed out by these authors, most of the mass is
ejected at slow speed in the pre-eruptive wind, while most of the kinetic
energy is supplied by the fast material with much lower mass loss.

\begin{figure*}
\hspace{-2.0cm}\includegraphics[width=10.5in,angle=0]{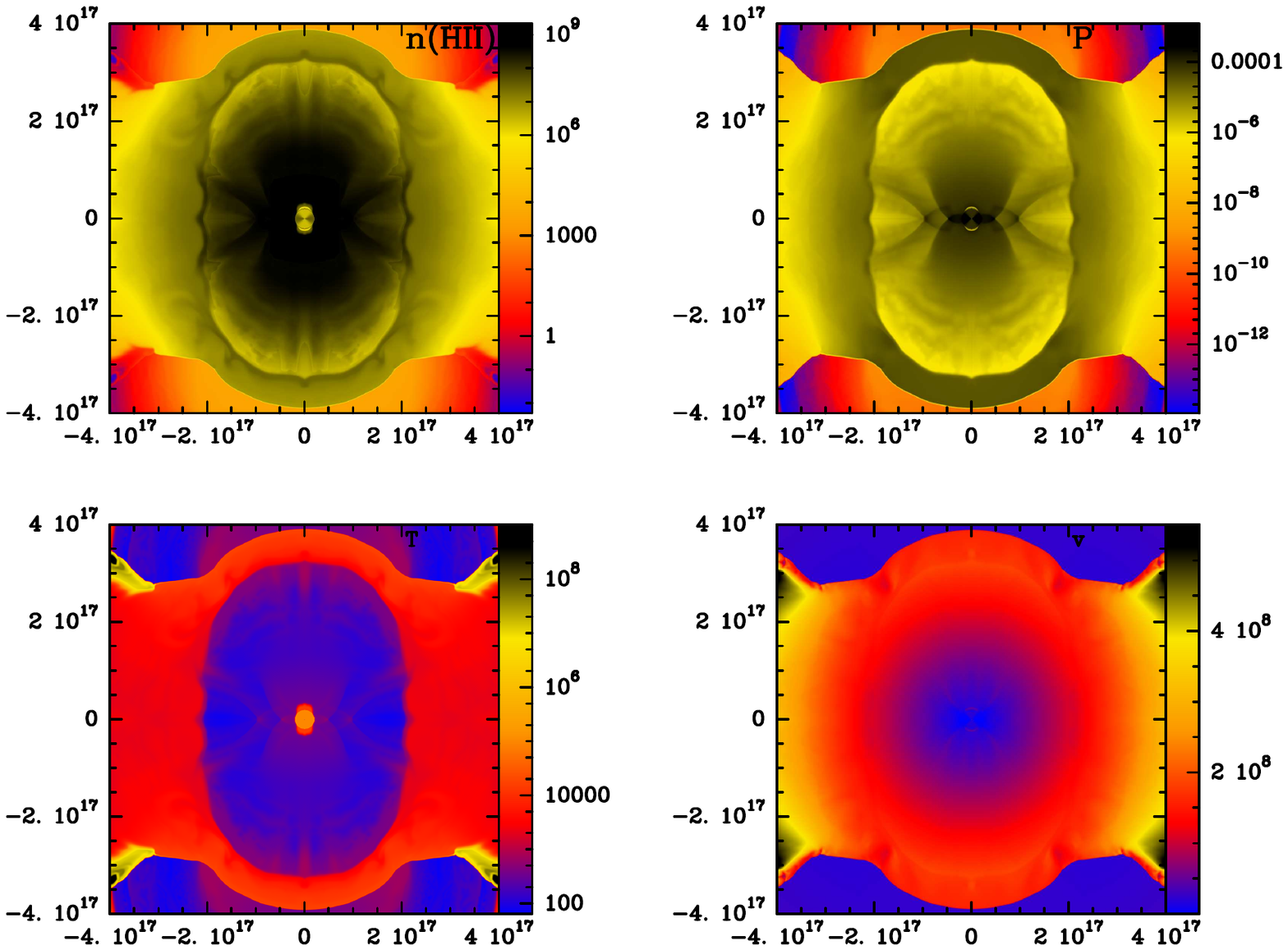}
\vspace{-4cm}
    \caption{Stratifications for Run B of density of ionized hydrogen (top left),
pressure (top right), temperature (bottom left) and velocity (bottom right)
computed 52 yr after the great eruption of $\eta$ Car. The bars at the right-side
of the panels are in units of cm$^{-3}$, dyn cm$^{-2}$, K, and cm s$^{-1}$,
respectively, and the computational domain axes are in units of cm. A further
description of the figure is given in the text.}
    \label{fig4}
\end{figure*}

In Figure 3, we show the density map (in logarithmic scale) obtained from
Run B at an evolution time $t =$ 52 yr since the explosive event. For this
model, we have assumed for the great eruption a wide-angle explosive outflow 
with a speed $v_1 = 10000$ km s$^{-1}$ and a speed $v_2 = 100$ km s$^{-1}$
with most of the mass expelled toward the symmetry axis,
such as it was proposed by Smith et al. (2018a). We note from the figure that
this model has an important impact on the morphology and kinematics of $\eta$
Car's nebulae. The assumed latitude-dependent flow parameters at injection are
not able to account for the shape of the large Homunculus, that is, the
outer expanding shell does not resemble its bipolar structure. Furthermore,
the shocked layer expands too quickly, and therefore reaches the current physical
size of $\sim 3.5 \times 10^{17}$ cm (in the polar direction) at an evolution
time $t= 52$ yr since the explosion, long before the estimated kinematical
age of $\sim$175 yr of the large Homunculus. On the other
hand, the resulting expansion speed of $\simeq$ 1800 km s$^{-1}$ at the poles
is much greater than the observed value of $\simeq$ 650 km s$^{-1}$\citep{smi2006}.
At the equator, the numerical simulation shows a physical size of the
nebula of $\pm$2.3 $\times$ 10$^{17}$ cm expanding at $\simeq$ 1200 km s$^{-1}$,
which are not consistent with the corresponding observed values
\citep[$\pm$3.1 $\times$ 10$^{16}$ cm and $\simeq$ 62 km s$^{-1}$;][]{smi2006}.
Besides, it is worth mentioning that important differences from our previous
models \citep{gon2010,gon2018}  are identified on the embedded structures.
In particular, the simulation shows the formation of tenuous
polar caps that may be related to the inner Homunculus, which results from
the interaction of the minor eruption with its post-outburst wind. Nonetheless,
these caps are located at a distance of $\pm$2.7
$\times$ 10$^{16}$ cm from the central source, that does not correspond
either to the current physical size of the {\it little} Homunculus of
$\pm$ 8 $\times$ 10$^{16}$ cm \citep{ish2003,smi2006}.

Figure 4 shows the density of ionized hydrogen (top left), pressure (top right),
temperature (bottom left) and velocity (bottom right) maps computed for the
interaction of the different outflows for Run B. The stratifications correspond
to an evolution time of $t =$ 52 yr after of the explosion phase. As described
above, the simulation shows that the outer shocked layer expands too fast reaching
the observed physical size long before the estimated age of 175 yr of the
large Homunculus. The simulation shows a temperature
of $T \simeq$ 1.6 $\times 10^4$ K behind the outer shock (in the polar direction),
while the mean temperature inside the nebula is $T \simeq$ 200 K. At the equator,
outside the nebula, we see the presence of material with density of ionized hydrogen
$n_{HII} \simeq 10^6$ cm$^{-3}$, and temperature of $T \simeq$ 3 $\times 10^3$ K, and at
higher latitudes, the density decreases to $n_{HII} \simeq 1$ cm$^{-3}$ and
the temperature increases to $T \simeq$ 2 $\times 10^7$ K. In addition, the model
predicts high speeds in the equatorial plane, with values increasing from $v
\simeq$ 1000 km s$^{-1}$ to $v \simeq$ 3000 km s$^{-1}$. At higher latitudes,
the low-density material expands at a higher speed of $v \simeq$ 5000 km s$^{-1}$.
Notwithstanding, no equatorial low-density features that may be related to the
equatorial skirt of $\eta$ Car nebula are observed.

\section{Conclusions}

In this article, new two-dimensional numerical models of the 1840s
great eruption of $\eta$ Car are presented. In contrast with our
previous models which assume this eruption as a single explosion \citep{gon2018},
in the new high-resolution hydrodynamical simulations, we have incorporated
the intermediate (500 km s$^{-1}$) and high-speed (10000 km s$^{-1}$) components 
proposed recently by \citet[][]{smi2018,smi2018b}. 
During this event, the energy injection was $\sim 10^{49}-10^{50}$ erg
in which the parameters of $\eta$ Car wind (ejection velocity and mass-loss
rate) may have drastically changed in a very short period of time ($\sim$ 1 yr).
Consequently, we have assumed here that the {\it large} Homunculus nebula may form
from the interaction of this explosion with a previous stage wind of
$\eta$ Car \citep[the CSM scenario proposed by][]{smi2013}. It is worth mentioning
that we have not addressed the inner mechanism that triggered the
explosive event. Potential processes are discussed in Smith et al. (2018a).

Adopting this scenario, our numerical simulation with an intermediate-speed
explosion (1000 km s$^{-1}$, Run A) was able to explain both the shape and kinematics
of the {\it large} Homunculus. In addition, the simulation predicts a kinematical
age of the nebula of $\sim$ 174 yr that is consistent with the value estimated
from observations ($\sim$ 176 yr). At this time of evolution, the predicted expansion
velocity - along the symmetry axis - of the external shocked layer is of 
$\simeq$ 600 km s$^{-1}$, which is very similar to the value of
$\simeq$ 650 km s$^{-1}$ of the eta Car's Homunculus observed by \citet{smi2006}.
Furthermore, this model also shows the formation of the internal nebula (the
{\it little} Homunculus) from the interaction of the minor eruption of the 1890s
with the post-outburst wind, which, as in our previous models \citep{gon2004,gon2010} 
developed Rayleigh-Taylor instabilities that resemble the observed spatial structures
in the polar caps \citep{smi2005}. Nonetheless, the high-speed equatorial features
that may be related to the equatorial skirt of $\eta$ Carinae are not produced.

On the other hand, the high-speed explosion numerical model (10000 km s$^{-1}$, Run B)
shows important differences both with regard to the morphology and dynamical evolution
of the {\it large} Homunculus. First, the simulation predicts an outer shocked
layer that does not resemble the bipolar structure of the nebula, and second,
this layer expands too quickly (at a speed of $\simeq$ 1800 km s$^{-1}$)
and, consequently, it reaches the observed physical size of the nebula ($\sim 3.2
\times 10^{17}$ cm) -along the symmetry axis- at a time of 52 yr after the explosion,
i.e. long before the estimated age of $\sim$ 176 yr of the {\it large} Homunculus.
Besides, the numerical simulation shows important differences in the embedded
structures. As expected, the model shows the formation of tenuous polar caps
from the interaction of the minor eruptive event with the post-out burst wind,
however, at this time of evolution, these structures are located at a distance
of $\pm\, 2.7 \times 10^{17}$ cm from the source, and therefore do not reproduce
the estimated extension of $\pm\,8 \times 10^{17}$ cm of the polar caps of the
{\it little} Homunculus. We should note that despite the differences in the parameters
of Runs A and B, if we had employed the same values for the density profile, the
general features obtained for the explosive model Run B would be essentially the
same.

A final remark is that, in this work, we have focused in the shape and dynamical
evolution of the {\it large} Homunculus assuming a high-speed, 10000 km s$^{-1}$
explosion for the 1840s great eruption of $\eta$ Car. One can ask whether an
explosive event with speeds as high as 10000 km s$^{-1}$ or larger is able to
account for the observed physical properties of $\eta$ Car's nebula. We conclude
from our numerical simulation that included this high-speed component (Run B)
that this model can not reproduce the morphology or the age estimated from
observations of the {\it large} Homunculus. On the other hand, a slower but
also brief event with a velocity of 1000 km s$^{-1}$ could nearly reproduce
the observations.

\begin{acknowledgements}
RFG acknowledges financial support for UNAM-PAPIIT grant IN 107120.
L.A.Z. acknowledges financial support from CONACyT-280775 and UNAM-PAPIIT
IN110618 grants. ACR, JC, and PFV acknowledge the financial support for
UNAM-PAPIIT grants IG 1000218 and IA10321. EMdGDP is grateful for the support
from the Brazilian agencies FAPESP (grant 2013/10559-5) and CNPq (grant
308643/2017-8). The authors thank the anonymous referee for her/his useful
comments and suggestions that improved the content of this work.
\end{acknowledgements}




\bibliographystyle{aa}

\begin{thebibliography}{}

\bibitem[Abraham et al.(2014)]{abr2014} Abraham, Z., Falceta-Gon{\c{c}}alves, D., \& Beaklini, P.~P.~B.\ 2014, \apj, 791, 95. doi:10.1088/0004-637X/791/2/95
\bibitem[Davidson \& Humphreys(1997)]{dav1997} Davidson, K., \& Humphreys, R.~M.\ 1997, \araa, 35, 1
\bibitem[Davidson et al.(2001)]{dav2001} Davidson, K., Smith, N., Gull, T.~R., et al.\ 2001, \aj, 121, 1569. doi:10.1086/319419
\bibitem[Humphreys et al.(1999)]{hum1999} Humphreys, R.~M., Davidson, K., \& Smith, N.\ 1999, \pasp, 111, 1124. doi:10.1086/316420
\bibitem[Icke(1988)]{ick1988} Icke, V.\ 1988, \aap, 202, 177
\bibitem[Ishibashi et al.(2003)]{ish2003} Ishibashi, K., Gull, T.~R., Davidson, K., et al.\ 2003, \aj, 125, 3222. doi:10.1086/375306
\bibitem[Gonz{\'a}lez(2018)]{gon2018} Gonz{\'a}lez, R.~F.\ 2018, \aap, 609, A69 
\bibitem[Gonz{\'a}lez et al.(2010)]{gon2010} Gonz{\'a}lez, R.~F., Villa, A.~M., G{\'o}mez, G.~C., et al.\ 2010, \mnras, 402, 1141. doi:10.1111/j.1365-2966.2009.15950.x
\bibitem[Gonz{\'a}lez et al.(2004a)]{gon2004} Gonz{\'a}lez, R.~F., de Gouveia Dal Pino, E.~M., Raga, A.~C., et al.\ 2004, \apj, 616, 976. doi:10.1086/425112
\bibitem[Gonz{\'a}lez et al.(2004b)]{gon2004b} Gonz{\'a}lez, R.~F., de Gouveia Dal Pino, E.~M., Raga, A.~C., et al.\ 2004, \apjl, 600, L59. doi:10.1086/381390
\bibitem[Kashi \& Soker(2009)]{kas2009} Kashi, A., \& Soker, N.\ 2009, \na, 14, 11 
\bibitem[Morris et al.(1999)]{mor1999} Morris, P.~W., Waters, L.~B.~F.~M., Barlow, M.~J., et al.\ 1999, \nat, 402, 502.
\bibitem[Smith et al.(2018a)]{smi2018} Smith, N., Andrews, J.~E., Rest, A., et al.\ 2018, \mnras, 480, 1466.
\bibitem[Smith et al.(2018b)]{smi2018b} Smith, N., Rest, A., Andrews, J.~E., et al.\ 2018, \mnras, 480, 1457.
\bibitem[Smith(2013)]{smi2013} Smith, N.\ 2013, \mnras, 429, 2366. doi:10.1093/mnras/sts508
\bibitem[Smith(2006)]{smi2006} Smith, N.\ 2006, \apj, 644, 1151
\bibitem[Smith(2005)]{smi2005} Smith, N.\ 2005, The Fate of the Most Massive Stars, 332, 307
\bibitem[Smith et al. (2003)]{smi2003} Smith N., Gehrz R.~D., Hinz P.~M., Hoffmann W.~F., Hora J.~L., Mamajek E.~E., Meyer M.~R., 2003, AJ, 125, 1458
\bibitem[Owocki \& Shaviv(2016)]{owo2016} Owocki, S.~P., \& Shaviv, N.~J.\ 2016, \mnras, 462, 345 
\bibitem[Portegies Zwart, \& van den Heuvel(2016)]{por2016} Portegies Zwart, S.~F., \& van den Heuvel, E.~P.~J.\ 2016, \mnras, 456, 3401.
\bibitem[Raga et al.(2000)]{rag2000} Raga, A.~C., Navarro-Gonz{\'a}lez, R., \& Villagr{\'a}n-Muniz, M.\ 2000, \rmxaa, 36, 67
\bibitem[Raga et al.(2001)]{rag2001} Raga, A. C., Villagran-Muniz, M. Navarro-Gonz\'alez, R., Masciadri,  E. 2001, RMxAA, 37, 87
\bibitem[Raga et al.(2002)]{rag2002} Raga, A. C., de Gouveia Dal Pino, E. M., Noriega-Crespo, A., Mininni, P. D., Vel\'azquez, P. F. 2002, A\&A, 392, 267
\bibitem[van Leer(1982)]{van1982} van Leer, B.\ 1982, Numerical Methods in Fluid Dynamics, 507. doi:10.1007/3-540-11948-5\_66
\bibitem[Vel\'azquez et al. (2001)]{vel2001} Vel\'azquez. P. F., Sobral, H., Raga, A. C., Villagr\'an-Muniz, M., Navarro-Gonz\'alez, R. 2001, RMxAA, 37, 87
\bibitem[Woosley(2017)]{woo2017} Woosley, S.~E.\ 2017, \apj, 836, 244

\end{thebibliography}

\end{document}